\documentclass[twocolumn]{aastex701}
\usepackage[varg]{txfonts}
\usepackage{comment}
\usepackage{graphicx}
\usepackage{txfonts}
\usepackage{multirow}
\usepackage{subcaption}
\usepackage{adjustbox}
\usepackage{amssymb}
\usepackage{xcolor}
\usepackage{colortbl}

\usepackage{textgreek}
\usepackage{xargs}
\usepackage{xcolor}
\usepackage{xspace}

\let\oldtextsigma\textsigma
\renewcommand{\textsigma}{\oldtextsigma\xspace}
\let\oldAA\AA
\renewcommand{\AA}{\text{\oldAA}\xspace}
\def\w80{\ensuremath{w_{80}}\xspace}

\newcommandx{\fluxdcgs}[1][1=-20]{$\times 10^{[#1]}$~erg~s$^{-1}$~cm$^{-2}$~\AA$^{-1}$\xspace}

\newcommand{\Halpha}{\text{H\textalpha}\xspace}
\newcommand{\Hbeta}{\text{H\textbeta}\xspace}
\newcommand{\Hgamma}{\text{H\textgamma}\xspace}
\newcommand{\Hdelta}{\text{H\textdelta}\xspace}

\newcommandx{\permittedEL}[6][1=O,2=III,3=,4=,5=,6=]{\text{{#1}\,{\sc{#2}}{#3}{#4}{#5}{#6}}\xspace}
\newcommandx{\semiforbiddenEL}[6][1=O,2=III,3=,4=,5=,6=]{\text{{#1}\,{\sc{#2}}]{#3}{#4}{#5}{#6}}\xspace}
\newcommandx{\forbiddenEL}[6][1=O,2=III,3=,4=,5=,6=]{\text{[{#1}\,{\sc{#2}}]{#3}{#4}{#5}{#6}}\xspace}

\newcommandx{\HI}{\permittedEL[H][i]}
\newcommandx{\HII}{\permittedEL[H][ii]}
\newcommandx{\HeI}{\permittedEL[He][i]}
\newcommandx{\HeIL}[1][1=3889]{\permittedEL[He][i][\,\textlambda][#1]}
\newcommandx{\HeIIL}[1][1=4686]{\permittedEL[He][ii][\,\textlambda][#1]}

\newcommandx{\OIIIL}[1][1=5007]{\forbiddenEL[O][iii][\textlambda][#1]}

\newcommandx{\NIL}[1]{\forbiddenEL[N][i][\textlambda][5200]}
\newcommand{\OI}{\permittedEL[O][i]}
\newcommandx{\OIL}[1][1=8446]{\permittedEL[O][i][\textlambda][#1]}
\newcommandx{\OILs}[1][1=8446]{\text{\textlambda {#1}}\xspace}
\newcommand{\OIall}{\permittedEL[O][i][\textlambda][][1304,7774,7790,8446]}

\newcommand{\OII}{\forbiddenEL[O][ii]}
\newcommandx{\OIIL}[1][1=3727]{\forbiddenEL[O][ii][\textlambda][#1]}

\newcommand{\NeIIIall}{\forbiddenEL[Ne][iii][\textlambda][\textlambda][][3869,3967]}

\newcommand{\NIV}{\semiforbiddenEL[N][iv]}
\newcommand{\NV}{\permittedEL[N][v]}
\newcommand{\NII}{\forbiddenEL[N][ii][\textlambda][6584]}
\newcommandx{\NIIL}[1][1=6583]{\forbiddenEL[N][ii][\textlambda][#1]}

\newcommandx{\CIV}{\permittedEL[C][iv]}

\newcommandx{\CaT}{\permittedEL[Ca][ii][\textlambda][][8498,8542,8662]}
\newcommandx{\CaL}[1][1=8498]{\permittedEL[Ca][ii][\textlambda][#1]}
\newcommandx{\Ca}{\permittedEL[Ca][ii]}

\newcommandx{\Fef}{\forbiddenEL[Fe][ii]}

\newcommandx{\Fe}{\permittedEL[Fe][ii]}
\newcommandx{\FeL}{\permittedEL[Fe][ii][\textlambda]}

\newcommandx{\Feopt}{\permittedEL[Fe][ii][\textlambda][\textlambda][5190,][5320]}

\newcommandx{\CIII}{\semiforbiddenEL[C][iii]}

\newcommandx{\MgII}{\permittedEL[Mg][ii]}

\newcommand{\SiII}{\permittedEL[Si][ii]}

\newcommand{\SII}{\forbiddenEL[S][ii]}

\graphicspath{{./}{figures/}}

\begin{document}

\title{A deep dive down the broad-line region: permitted \OI, \Ca and \Fe emission in an AGN Little Red Dot at $z=5.3$}


\author[orcid=0000-0002-9909-3491,sname='RT']{Roberta Tripodi}
\affiliation{INAF - Osservatorio Astronomico di Roma, Via Frascati 33, I-00078 Monte Porzio Catone, Italy}
\affiliation{University of Ljubljana FMF, Jadranska 19, 1000 Ljubljana, Slovenia}
\affiliation{IFPU - Institute for Fundamental Physics of the Universe, via Beirut 2, I-34151 Trieste, Italy}
\email[show]{roberta.tripodi@inaf.it}  

\author[orcid=0000-0001-5984-03952]{Maruša Bradač} 
\affiliation{University of Ljubljana FMF, Jadranska 19, 1000 Ljubljana, Slovenia}
\email{fakeemail2@google.com}

\author[orcid=0000-0003-2388-8172]{Francesco D'Eugenio} 
\affiliation{Kavli Institute for Cosmology, University of Cambridge, Madingley Road, Cambridge CB3 0HA, UK}
\affiliation{Cavendish Laboratory - Astrophysics Group, University of Cambridge, 19 JJ Thomson Avenue, Cambridge, CB3 0HE, UK}
\email{fakeemail2@google.com}

\author[orcid=0000-0003-3243-9969]{Nicholas Martis} 
\affiliation{University of Ljubljana FMF, Jadranska 19, 1000 Ljubljana, Slovenia}
\email{fakeemail2@google.com}

\author[orcid=0009-0009-4388-898X]{Gregor Rihtaršič} 
\affiliation{University of Ljubljana FMF, Jadranska 19, 1000 Ljubljana, Slovenia}
\email{fakeemail2@google.com}

\author[orcid=0000-0002-4201-7367]{Chris Willott} 
\affiliation{NRC Herzberg, 5071 West Saanich Rd, Victoria, BC V9E 2E7, Canada}
\email{fakeemail2@google.com}

\author[orcid=0000-0001-8940-6768]{Laura Pentericci} 
\affiliation{INAF - Osservatorio Astronomico di Roma, Via Frascati 33, I-00078 Monte Porzio Catone, Italy}
\email{fakeemail2@google.com}

\author[]{Bianca Moreschini}
\affiliation{Dipartimento di Fisica e Astronomia, Università di Firenze, Via G. Sansone 1, 50019 Sesto F.no (Firenze), Italy}
\affiliation{INAF-Osservatorio Astrofisico di Arcetri, Largo E. Fermi 5, 50125 Firenze, Italy}
\email{fakeemail2@google.com}

\author[0000-0003-0144-4052]{Maxim Markevitch}
\affiliation{NASA / Goddard Space Flight Center, Greenbelt, MD 20771, USA}
\email{fakeemail2@google.com}

\author[orcid=0000-0003-3983-5438]{Yoshihisa Asada}
\affiliation{Waseda Research Institute for Science and Engineering, Faculty of Science and Engineering, Waseda University, 3-4-1 Okubo, Shinjuku, Tokyo 169-8555, Japan}
\email{fakeemail2@google.com}

\author[orcid=0000-0003-2536-1614]{Antonello Calabrò} 
\affiliation{INAF - Osservatorio Astronomico di Roma, Via Frascati 33, I-00078 Monte Porzio Catone, Italy}
\email{fakeemail2@google.com}

\author[orcid=0000-0001-8325-1742]{Guillaume Desprez} 
\affiliation{Kapteyn Astronomical Institute, University of Groningen, P.O. Box 800, 9700AV Groningen, The Netherlands}
\email{fakeemail2@google.com}

\author[]{Giordano Felicioni} 
\affiliation{University of Ljubljana FMF, Jadranska 19, 1000 Ljubljana, Slovenia}
\email{fakeemail2@google.com}

\author[orcid=0000-0001-9293-4449]{Gaia Gaspar} 
\affiliation{Department of Astronomy and Physics and Institute for Computational Astrophysics, Saint Mary's University, 923 Robie Street, Halifax, B3H 3C3, Nova Scotia}
\email{fakeemail2@google.com}

\author[orcid=0000-0002-0933-8601]{Anthony H. Gonzalez}
\affiliation{Department of Astronomy, University of Florida, Bryant Space Science Center, Gainesville, FL 32611, USA}
\email{fakeemail2@google.com}

\author[orcid=0000-0001-9414-6382]{Anishya Harshan} 
\affiliation{University of Ljubljana FMF, Jadranska 19, 1000 Ljubljana, Slovenia}
\email{fakeemail2@google.com}

\author[orcid=0000-0002-1660-9502]{Xihan Ji} 
\affiliation{Kavli Institute for Cosmology, University of Cambridge, Madingley Road, Cambridge CB3 0HA, UK}
\affiliation{Cavendish Laboratory - Astrophysics Group, University of Cambridge, 19 JJ Thomson Avenue, Cambridge, CB3 0HE, UK}
\email{fakeemail2@google.com}

\author[orcid=0009-0000-2101-1938]{Jon Judež} 
\affiliation{University of Ljubljana FMF, Jadranska 19, 1000 Ljubljana, Slovenia}
\email{fakeemail2@google.com}

\author[orcid=0000-0002-1428-7036]{Brian C. Lemaux}
\affiliation{Gemini Observatory, NSF NOIRLab, 670 N. A'ohoku Place, Hilo, Hawai'i, 96720, USA}
\affiliation{Department of Physics and Astronomy, University of California, Davis, One Shields Ave., Davis, CA 95616, USA}
\email{fakeemail2@google.com}

\author[orcid=0000-0002-9889-4238]{Alessandro Marconi}
\affiliation{Dipartimento di Fisica e Astronomia, Università di Firenze, Via G. Sansone 1, 50019 Sesto F.no (Firenze), Italy}
\affiliation{INAF-Osservatorio Astrofisico di Arcetri, Largo E. Fermi 5, 50125 Firenze, Italy}
\email{fakeemail2@google.com}

\author[orcid=0000-0002-5694-6124]{Vladan Markov} 
\affiliation{University of Ljubljana FMF, Jadranska 19, 1000 Ljubljana, Slovenia}
\email{fakeemail2@google.com}

\author[orcid=0000-0001-8115-5845]{Rosa M. Merida} 
\affiliation{Department of Astronomy and Physics and Institute for Computational Astrophysics, Saint Mary's University, 923 Robie Street, Halifax, B3H 3C3, Nova Scotia}
\email{fakeemail2@google.com}

\author[orcid=0000-0002-8951-4408]{Lorenzo Napolitano} 
\affiliation{INAF - Osservatorio Astronomico di Roma, Via Frascati 33, I-00078 Monte Porzio Catone, Italy}
\email{fakeemail2@google.com}

\author[]{Gaël Noirot} 
\affiliation{Space Telescope Science Institute, 3700 San Martin Drive, Baltimore, Maryland 21218, USA}
\email{fakeemail2@google.com}

\author[orcid=0000-0002-9729-3721]{Massimiliano Parente}
\affiliation{INAF - Observatory of Trieste, Via G.B. Tiepolo 11, 34131, Trieste, Italy}
\email{fakeemail2@google.com}

\author[orcid=0000-0002-8040-6785]{Annika H. G. Peter}
\affiliation{Department of Physics, Department of Astronomy, and CCAPP, The Ohio State University}
\email{fakeemail2@google.com}

\author[orcid=0000-0002-6265-2675]{Luke Robbins} 
\affiliation{Department of Physics and Astronomy, Tufts University, 574 Boston Avenue, Suite 304, Medford, MA 02155, USA}
\email{fakeemail2@google.com}

\author[0000-0002-0086-0524]{Andrew Robertson}
\affiliation{Carnegie Observatories, 813 Santa Barbara Street, Pasadena, CA 91101, USA}
\email{fakeemail2@google.com}

\author[orcid=0000-0001-8830-2166]{Ghassan T. E. Sarrouh} 
\affiliation{Department of Physics and Astronomy, York University, 4700 Keele St. Toronto, Ontario, M3J 1P3, Canada}
\email{fakeemail2@google.com}

\author[orcid=0000-0002-7712-7857]{Marcin Sawicki} 
\affiliation{Department of Astronomy and Physics and Institute for Computational Astrophysics, Saint Mary's University, 923 Robie Street, Halifax, B3H 3C3, Nova Scotia}
\email{fakeemail2@google.com}

\begin{abstract}
    We present a spectroscopic analysis of a broad-line active galactic nucleus (AGN) selected as little red dot at $z = 5.3$ behind the Bullet cluster (Bz5.3), based on \textit{JWST}/NIRCam and NIRSpec data. The detection of strong \Fe, \OI, and \Ca triplet emission lines, along with the evidence of broad Balmer lines, provides strong evidence of a broad-line region (BLR) and an accreting supermassive black hole. Notably, we report the first detection of the $\lambda1304$ bump (i.e., blend of \OIL[1304] and \SiII) at high redshift, a feature commonly seen in local AGNs  but not yet reported in the early Universe. The \OIL[1304]/$\lambda8446$ photon ratio provides an independent measurement of dust attenuation in galaxies. In Bz5.3, this ratio is highly suppressed (0.1--0.3), implying significant internal dust extinction, with estimated dust attenuation $A_V \sim 0.4$--$1.0$. We identify Ly$\beta$ fluorescence as the dominant excitation mechanism of the low-ionization lines, with additional contributions from collisional excitation. High \OIL[8446] equivalent width and weak \OIL[7774] support this interpretation. 
    The detection of iron emission, whether from broad permitted or narrow forbidden lines, supports the presence of a stratified BLR, as also recently proposed in local LRDs. Photoionization modeling of \OIL[8446] and \Ca further suggests the coexistence of multiple gas phases with distinct densities and ionization states, highlighting the complexity of the BLR.
    Bz5.3 thus offers a rare window into early AGN activity and BLR physics at early times.

\end{abstract}


\section{Introduction}
\label{sec:intro}

The Broad Line Region (BLR) --- a compact, high-density ($n_{\rm h}\sim 10^9$--$10^{10}~\rm cm^{-3}$) and high-velocity (FWHM $\sim 1000$--$10000~\rm km~s^{-1}$) region within $\lesssim 1$~pc of a supermassive black hole (SMBH) --- produces the broad emission lines observed in active galactic nuclei (AGNs). These lines, including Balmer lines, \MgII, \CIV, \Fe, \OI, and the \Ca triplet, provide key insights into the structure, kinematics, and chemical composition of the gas surrounding SMBHs \citep[see][]{netzer2013}. Among them, low-ionization lines (LILs) like \Fe, \OI, and \CaT are especially valuable for probing the outer BLR.

Understanding the excitation and distribution of UV--optical iron emission is crucial, as \Fe is a major coolant of the BLR and reflects the gas energy budget \citep{wills1985, matsuoka2007}. Its emission also traces chemical enrichment across cosmic time \citep{hamann1999, hamann1993, yoshii1998}. The main excitation mechanisms are Ly$\alpha$ fluorescence and collisional excitation \citep{grandi1980, rodriguez2002}, with near-IR \Fe lines especially linked to the former. Thanks to studies using reverberation mapping, \Fe is found to arise in the outer BLR, where its spectral profile and equivalent width encode the physical conditions \citep{kovacevic2010, baldwin2004}. Interestingly, UV \Fe features have recently been reported in two AGN ``Little Red Dots'' (LRDs), A2744-45924 at $z=4.47$ \citep{labbe2024} and CAPERS-LRD-z9 at $z=9.288$ \citep{taylor2025}, directly linking them to BLR origin. It is still unknown whether these UV features are widespread
but weak, or if they emerge in certain LRDs due to specific BH properties or geometric effects such as inclination or the distribution of dust.
In contrast, optical \Fe lines are typically weak in JWST-selected broad-line AGN, possibly due to the reduced metallicity of their broad line region  \citep{trefoloni2024}. Despite the large number of studies of \Fe in AGNs \citep[e.g., ][]{marziani2013,netzer1983,wills1985,sigut1998,sigut2003,verner1999,verner2003,verner2004,baldwin2004,tsuzuki2006,rudy2000, dong2009, kovacevic2025,labbe2024,trefoloni2024, ferland2020}, interpreting the blended \Fe pseudo-continuum remains challenging given the complex energy levels of \Fe and the mixed excitation mechanisms. In detail, the \Fe emission is usually modeled by adopting templates generated by different photo-ionization models. However, models exhibit a high level of degeneracy when compared to the observed data, especially when the sensitivity and the spectral resolution of the data are low.

Complementary diagnostics come from simpler atoms thought to trace similar BLR regions and excitation conditions. Neutral oxygen (\OI) and singly-ionized calcium (\Ca) often correlate with \Fe in strength and width \citep{rodriguez2002, joly1987, matsuoka2007,matsuoka2008,marziani2013,ferland1989,persson1988}, reflecting their comparable ionization potentials (16.2, 13.6, and 11.9~eV for \Fe, \OI, and \Ca, respectively). The \OIL[8446] line lacks a narrow component and is associated with Ly$\beta$ fluorescence \citep{grandi1980, rudy1989}. Emission at \OIL[1304], \OIL[8446], and \OIL[11287] is expected in a 1:1:1 photon ratio, with deviations indicating alternative mechanisms like recombination or collisional excitation \citep{laor1997, rodriguez2002, matsuoka2005, matsuoka2007, matsuoka2008}. The \OIL[1304] line is usually found blended with the \SiII line, which constitutes the so-called \OILs [1304] bump. Additional \OI lines such as \OIL[7774] and \OIL[7990] help refine the dominant processes.

\Ca emission arises primarily from the near-IR triplet (\CaT), emitted from neutral, optically thick gas shielded from hard radiation \citep{ferland1989, joly1989}. Comparisons between \CaT, \OI, and \Fe emission, both in flux and kinematics, constrain the gas density, ionization, and geometry of the BLR \citep{matsuoka2007, matsuoka2008}.

\citet{matsuoka2008} found that these diagnostics in low-redshift quasars ($z<1$) show little redshift or luminosity dependence, suggesting similar gas densities and ionizing fields, possibly mixed with dust.

Until now, such studies have been largely limited to $z<1$ due to sensitivity and resolution limits. Only a few exceptions exist at higher redshifts: 4 quasars at $z\sim 2.1$ with weak \OIL[1304] detection \citep{espey1989}, a $z\sim2.26$ ``Rosetta Stone'' quasar with \Ca emission \citep{juodzbalis2024}, and the $z\sim4.5$ LRD A2744-45924 with \OI, \CaT.  After the discovery of the \OILs[1304] bump in Bz5.3, \OIL[1304] has been detected in  CANUCS-LRD-z8.6, an LRD at z=8.6 \citep{morishita2025}.

In this Letter, we present the discovery of Bullet-LRD-z5.3 (hereafter Bz5.3), an AGN host at $z_{\rm spec} = 5.2907\pm 0.0002$, magnified by a factor $\mu=3.7_{-0.1}^{+0.2}$. Using \textit{JWST}/NIRSpec prism spectroscopy, we report the first detection of the \OILs[1304] bump at $z > 5$, along with strong \OIL[8446], \CaT, and rest-UV/optical \Fe emission. These observations offer a rare opportunity to probe BLR structure and excitation at early cosmic times. Data are presented in Sect.~\ref{sec:data}, results in Sect.~\ref{sec:results}, discussion in Sect.~\ref{sec:disc}, and conclusions in Sect.~\ref{sec:concl}.

Throughout the paper we adopt the $\Lambda$CDM cosmology from \citet{planck2018}: $H_0=67.4 ~\rm km ~s^{-1} ~Mpc^{-1}$, $\Omega_m=0.315$, and $\Omega_{\Lambda}=0.685$. Thus, the angular scale is $6.238 ~\rm kpc/arcsec$ at $z=5.2907$.

\begin{figure*}
    \centering
    \includegraphics[width=0.9\linewidth]{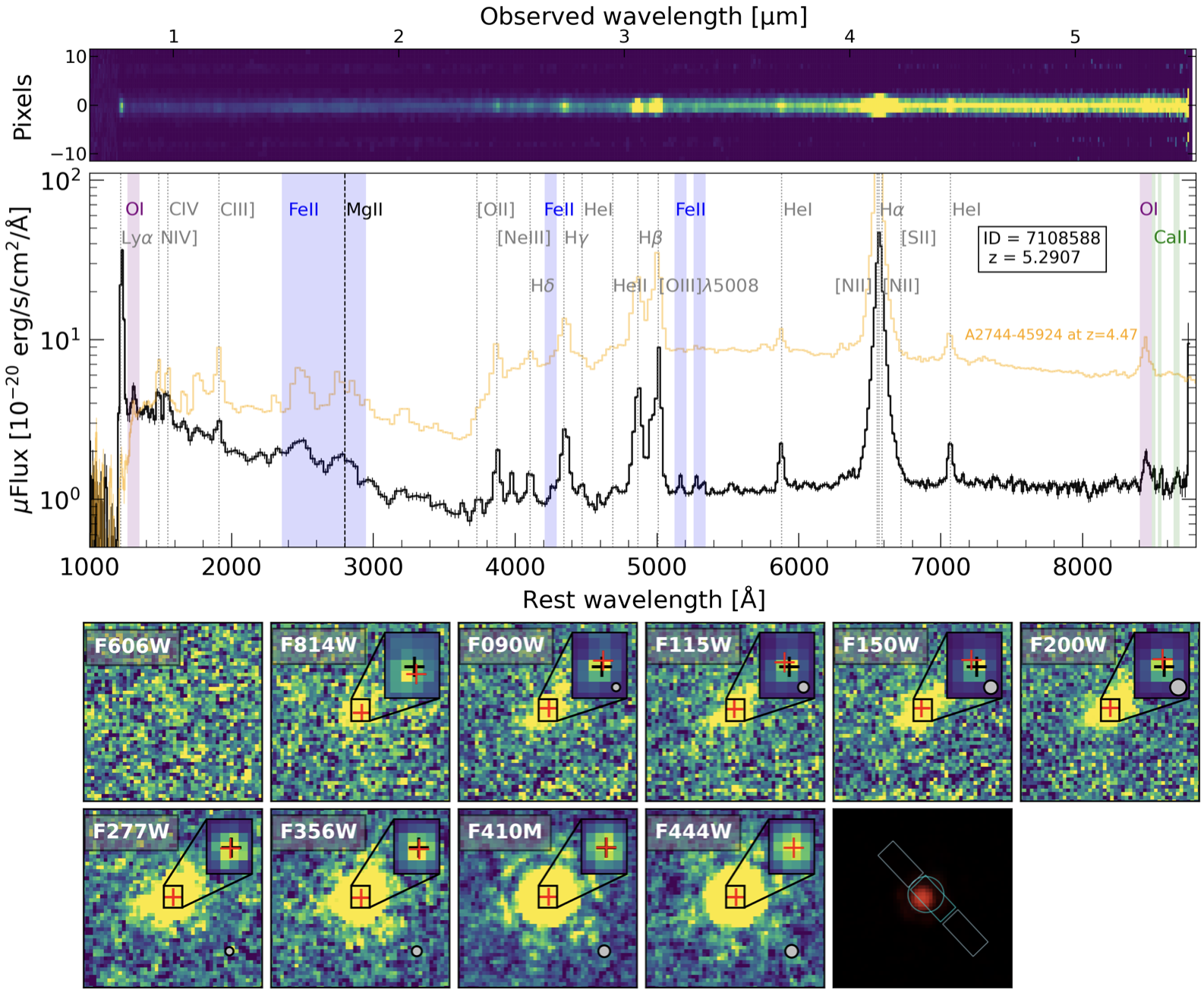}
    \caption{HST/ACS and NIRCam photometry, and NIRSpec spectra of Bz5.3. Top panels: 2D and 1D NIRSpec spectra. Fluxes are not corrected for magnification. Lines of interest for this work are highlighted by colored vertical bars. Other detected emission lines are marked with dashed gray vertical lines. The y-axis scale of the 1D spectrum is logarithmic. For comparison the spectrum of A2744-45924 at $z=4.47$ is shown in light orange. Bottom panels: NIRCam cut-out photometry in different filters, specified at the top left corner of each panel, and RGB image made combining the F444W, F356W, and F277W for red, F150W and F200W for green, and F090W and F115W filters for blue. Cutouts images are 2"$\times$2" in size. In each panel, there is an inset zooming-in on the center region to show the flux centroid of the filter (red cross), compared to the flux centroid in F444W (black cross). The PSF size of each NIRCam filter is shown as a gray circle in the bottom right corner of each panel (or zoom-in inset).}
    \label{fig:phot-spec}
\end{figure*}

\begin{figure*}
    \centering
    \includegraphics[width=0.98\linewidth]{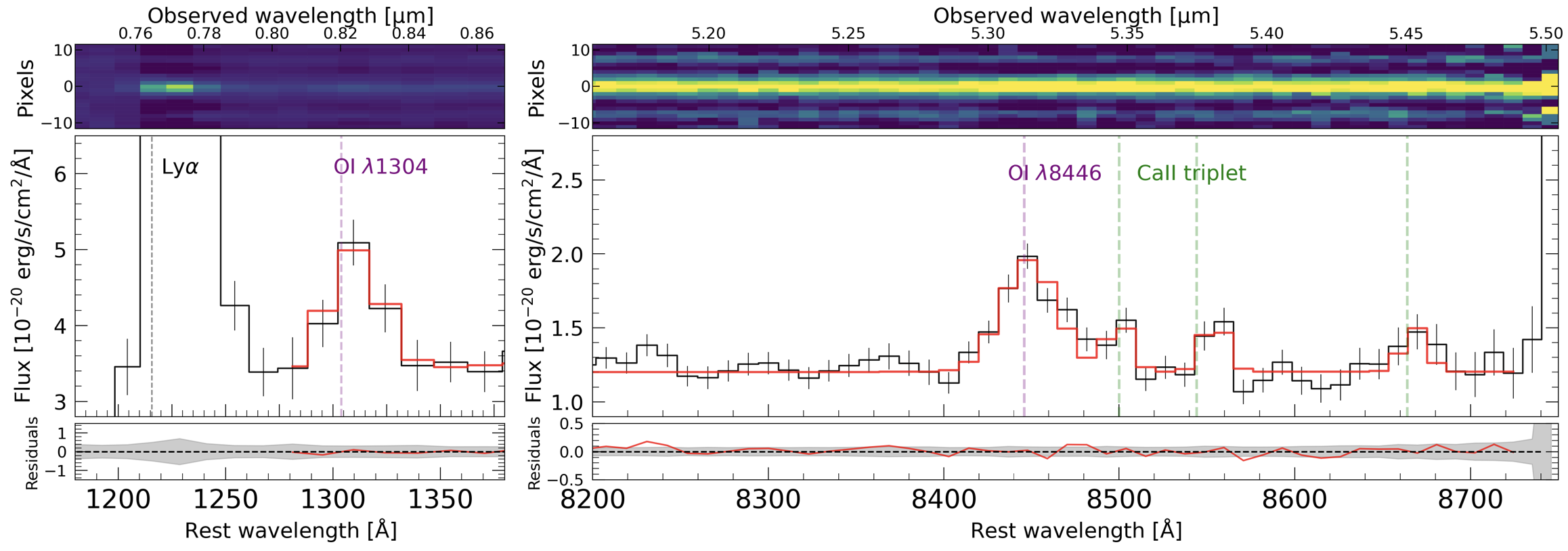}
    \caption{Zoom-in of the 2D and 1D spectra of Bz5.3. The observed spectrum is shown in black, and the best-fitting model is shown in red. Residuals are shown at the bottom of each panel, where the shaded gray area represents the 1$\sigma$ noise level. Left panel: zoom on the \OILs[1304] bump. The vertical pink line is at the laboratory rest wavelength of the \OIL[1304] emission line. The vertical dashed black line marks the Ly$\alpha$ line. Right panel: zoom-in on the \OIL[8446] emission line and on the \CaT triplet. Vertical lines are set at laboratory rest wavelength for these emission lines. The scale of the y-axis is linear.}
    \label{fig:zoom-spec}
\end{figure*}

\begin{figure*}
    \centering
    \includegraphics[width=0.98\linewidth]{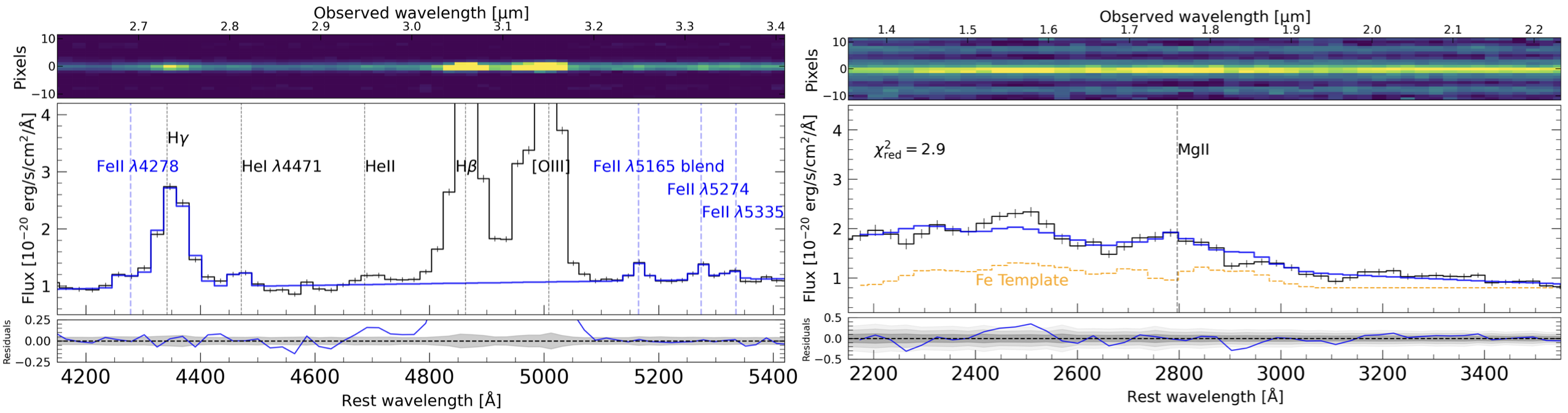}
    \caption{Zoom-in of the 2D and 1D spectra of Bz5.3. The observed spectrum is shown in black, while the best-fitting model is shown in blue. Residuals are shown at the bottom of each panel, where the shaded dark gray area represents the 1$\sigma$ noise level. Top panel: zoom on the \Fe emission in the optical. The vertical lines are at the laboratory rest wavelength of the reported emission line. Bottom panel: zoom on the \Fe complex in the UV and \MgII emission line. 2 and 3$\sigma$ noise levels are reported in the residual as shaded lighter gray area.}
    \label{fig:zoom-spec-Fe2}
\end{figure*}

\section{Data}
\label{sec:data}

\addtocounter{footnote}{-18}

The Bullet cluster was observed with NIRCam \citep{rieke2023} in \textit{JWST} Cycle 3 in January 2025 (GO program 4598, PI Bradač). We obtained imaging in eight filters, F090W, F115W, F150W, F200W, F277W, F356W, F410M and F444W with exposure times of $\sim 6.4$ ks each. In addition we use the archival HST/ACS observations in F606W, F775W and F850LP filters (programs 10200 and 10863, PI Jones \& Gonzalez) and F814W filter (program 11099, PI Bradač). The images were processed with a modified version of the stage 1 and stage 2 JWST pipeline, the Canadian NIRISS Unbiased Cluster Survey (CANUCS) data reduction pipeline \citep[full details in][]{sarrouh2025}, before being drizzled onto a 40 milliarcsec pixel grid using the \texttt{grizli} code \citep{brammer2019}. Bright cluster and foreground galaxies as well as intracluster light are removed from the images following the procedure outlined in \citet{martis2024}. Photometry is performed on the subtracted, psf-matched images convolved to the F444W filter resolution. 

Bz5.3 was photometrically selected as an LRD using the NIRCam data, following standard color and compactness criteria \citep{kokorev2024}.  Specifically, F115W-F150W$<0.8$, F200W-F277W$>0.7$, F200W-F356W$>1.0$, and the compactness in F444W is $<1.7$. Bz5.3 was followed up with NIRSpec/prism ($R\sim 30-400$) with the Micro-Shutter Assembly (MSA) as part of the same GO program 4598. NIRSpec/prism data have also been reduced and the redshift of the source has been estimated following the procedure described in \citet{sarrouh2025}. From the 2D spectral reduction, a 1D spectrum is extracted using an optimal extraction based on the source spatial profile. We correct for slit losses by scaling the 1D spectrum to the observed photometry. Specifically, we perform photometry in the region covered by the MSA slit, then fit a Chebyeshev polynomial to the ratio of the observed photometry and synthetic photometry obtained from convolving the spectrum within each filter. The 1D spectrum is then multiplied by the derived polynomial. 

To estimate magnification due to lensing, a strong lensing model was derived using the parametric lens modeling tool \texttt{Lenstool} \citep{Kneib96,Jullo07,Jullo09} and is constrained with a catalogue of 141 multiple images from 28 background galaxies with spectroscopic redshifts. The catalogue of multiple images was derived using JWST data. It includes multiple image candidates included in \citet{cha25} lens model, harnessing NIRCam imaging, and adds additional systems and  NIRSpec spectroscopic redshifts for 22 galaxies. The model includes large-scale dark matter halos, cluster member subhalos, cluster gas, included as a mass map from Chandra X-ray observations, and group-scale substructure halos surrounding the cluster.  The derived magnification of Bz5.3 is $\mu=3.7_{-0.1}^{+0.2}$ . 
For the discussion of the lens model and the catalogues of strong lensing constraints see Rihtaršič et al, in prep. Unless otherwise specified, quantities are corrected for magnification.  Magnification uncertainties are added in quadrature to the statistical uncertainties in fluxes, and introduce -on average- a systematic error of 0.03 dex in the flux and luminosities measurements. Magnification uncertainties are typically subdominant with respect to the statistical uncertainties on the fluxes. Reported errors on black hole mass estimates include propagation of magnification uncertainties.

An inspection of the available Chandra 500 ks data targeting the Bullet Cluster \citep{clowe2006} reveals no significant X-ray point source at the position of Bz5.3, consistent with the lack of X-ray detections reported in other LRD samples \citep{ananna2024, yue2024, sacchi2025}. 
A $3\sigma$ upper limit on the point source unabsorbed X-ray flux in the 0.5--2 kev
observer band is $1.2\times 10^{-7}$ phot\,cm$^{-2}$\,s$^{-1}$, with the
uncertainty dominated by the relatively bright diffuse emission of the galaxy
cluster plasma at this position. At $z=5.29$ and assuming $\mu=$3.7, this corresponds to the rest-frame 3--12 keV intrinsic luminosity of $L_X<1.6\times 10^{43}$ erg\,s$^{-1}$.

\section{Analysis and results}
\label{sec:results}

Figure~\ref{fig:phot-spec} show the 2D and 1D spectra of Bz5.3, along with all the available NIRCam photometry of the source. Bz5.3 is a point source in F444W and F410M, while shows extended emission in all the other filters (see App. \ref{app:galfit} for more details). Based on the flux centroids, we find no significant spatial variation ($\lesssim 0.04''$) between the emitting region in the bluest UV (F814w) and in near UV (F150W).

The blue UV slope ($\beta_{\rm UV, spec}=-1.6 \pm 0.1$), red optical slope ($\beta_{\rm opt, spec}=0.66 \pm 0.09$), and compactness in F444W together classify Bz5.3 as an LRD, in compliance with the standard criteria reported in \citet{kokorev2024} and \citet{kocevski2025}.

Several clear AGN signatures and a prominent Balmer break are evident upon visual inspection of the spectra. Notably, we detect Ly$\alpha$, \OIL[1304], \NIV, \CIV, \CIII, \MgII and \Fe, \OII, \NeIIIall, \Hdelta, \FeL[4278], \Hgamma, narrow \Hbeta, \HeIL[4471], \OIIIL, \FeL[5165]$\lambda 5274$$\lambda 5335$, \HeIL[5877], \NII, narrow \Halpha, \SII, \HeIL[7067], \OIL, \CaT, emission lines, as well as broad \Halpha emission with a full width at half maximum (FWHM) of $4600 \pm 50~\rm km~s^{-1}$, and broad \Hbeta emission with FWHM~$=5100\pm 500~\rm km~s^{-1}$. Using the virial relation from \citet{reines2015}\footnote{\citet{reines2015} assumed a mean scale factor of $f=4.3$ \citep{grier2013}.}, this broad \Halpha line implies a black hole (BH) mass of $\log(M_{\rm BH}/\rm M_\odot) = 8.4 \pm 0.3$. This implies an Eddington ratio of $\lambda_{\rm Edd}=0.008_{-0.005}^{+0.016}$, having derived the bolometric luminosity $\log(L_{\rm bol}/\rm erg ~s^{-1})=44.4 \pm 0.5$ from the continuum luminosity at 3000 \AA, adopting the bolometric correction from \citet{richards2009}. Analogously, using the virial relations from \citet{greene2005} and \citet{vestergaard2006}\footnote{\citet{greene2005} assumed a scale factor of $f=0.75$ \citep{kaspi2000}, while \citet{vestergaard2006} assumed $f=5.5 \pm 1.9$ \citep{onken2004}.} for the broad \Hbeta line and having $L_{\rm \Hbeta, broad}=7.8^{+2.4}_{-1.9} \times 10^{41} \rm ~erg/s$ from our best-fit model, we derive a BH mass of $\log(M_{\rm BH}/\rm M_\odot) = 7.9 \pm 0.3$, which is fully consistent with the one from \Halpha.

In this work, however, we focus on the discussion of the BLR properties as derived from the \OI, \Ca, and \Fe emission lines. Therefore, in this section, we describe the fitting methodology and present the results for these lines only. A complete analysis and discussion of the other source's properties is deferred to a following and more extensive work (Tripodi et al., in prep.).

\subsection{Continuum and line fitting}
\label{sec:fit}

The underlying continuum of the \OILs[1304] bump and of \OIL + \CaT is modeled separately using a linear relation\footnote{The linear relation provided a better model for the underlying continuum of the \OILs[1304] bump than a power-law given that the available wavelength range of continuum emission is small, since the bump lies close to the Ly$\alpha$ and \NV emission.} and power-law, respectively, both with free slope and normalization. Each emission line is modeled as a single Gaussian.  

The FWHM is fixed to be the same for each line in the \CaT triplet. The relative intrinsic ratio of the triplet is found to be 1:9:5 in an optically thin regime, and flattens to 1:1:1 for optically thick media, as derived from analysis of stars \citep{guarcello2014}. \citet{rudy2000} found that, in the AGN I~Zw~1, this ratio is 1:1.15:0.83 probably due to different gas conditions w.r.t a simple optically thick homogeneous medium, and this assumption has been adopted also in the analysis of quasars by \citet{matsuoka2008}. Given the lack of precise prescriptions in this regard, we choose to leave the flux relative intensities free and see whether our results align, or not, with the above mentioned values.

As mentioned in Sect. \ref{sec:intro}, modeling the \Fe pseudo-continuum is particularly challenging, and empirical templates from low-redshift well-studied sources are often used to fit the observed data \citep[e.g.][]{labbe2024}. In particular, we adopted the template either from \citet{vestergaard2001} or from \citet{tsuzuki2006} to model the \Fe signatures in the UV, which are also blended with the \MgII emission. Alternatively, we also produced \Fe templates from tailored photo-ionization models (see Appendix \ref{app:fe-fit} for details). The \MgII line is fitted with a single Gaussian, and the underlying continuum with a power-law relation as described above. The template has been convolved with the resolution of the instrument. 

The \Fe emission lines detected in the optical, \FeL[4278], \FeL[5165], \FeL[5274] and \FeL[5335], are modeled as a single Gaussian each, and the underlying continuum as a power law. We do not adopt empirical templates since some of the observed intensities of optical \Fe lines are not as theoretically expected \citep[see Sect .\ref{sec:disc} and ][]{kovacevic2025}. The FWHM of all these transitions is fixed to be the same, since they are thought to arise from the same region. \FeL[4278] is blended with \Hgamma + \OIIIL[4364] and \HeIL[4471], therefore we also model \Hgamma+\OIIIL[4364] and \HeIL[4471] as two independent Gaussians. Other emission lines at $4200 \AA< \lambda_{\rm rest}< 5400 \AA$ have been masked. The results of the fit are shown in Figure~\ref{fig:zoom-spec} for the \OI and \Ca complexes, and in Figure~\ref{fig:zoom-spec-Fe2} for \Fe and \MgII.

Details on the specific fitting procedure used and the derivation of the measured quantities can be found in Appendix \ref{app:fit}. Flux, FWHM and EW of all the fitted lines are reported in Table \ref{tab:results}. FWHMs are corrected for instrumental broadening (see Appendix \ref{app:fit} for details).

\subsection{Constitution of the \OILs[1304] Bump}
\label{sec:bump}

The \OILs[1304] bump typically seen in AGN spectra is the result from a blending of an \OI triplet at \OILs[1302.17], \OILs[1304.86], and \OILs[1306.03] (named so far as \OIL[1304]), and a \SiII doublet at \OILs[1304.37] and \OILs[1309.27] (hereafter just \SiII). The blending is usually very strong in most quasars, preventing a reliable estimate of the relative contributions of \OI and \SiII to the blend. However, \citet{laor1997} were able to estimate the \OI and \SiII fractions in the blend for I~Zw~1, given that the \SiII was clearly detected. They found that 50\% (56\%) of the bump flux is due to \OI in the optically thin (thick) \SiII doublet case. Moreover, \citet{rodriguez2002} estimated that in three narrow-line Seyfert 1 galaxies the average proportion of the \OI flux is 75\%, while \citet{matsuoka2007} reported a larger range of \OI fractions in 5 quasars at $0.15<z<1.1$, ranging from 20\% to 98\%.  \citet{matsuoka2007} also presented some theoretical considerations regarding the possible strengthening of the \SiII line: in particular, \SiII flux can be enhanced in the large optically emitting cloud (LOC) scenario or in a gas with unusually large micro-turbulent velocities \citep[as also suggested by][]{bottorff2000}. The latter scenario, however, implies a significant suppression in the Ly$\beta$ fluorescence rate, which is not our case as discussed in Sect. \ref{sec:ext-mec}. 

The spectral resolution of our data prevents us from resolving the bump; therefore, we fit the bump with a single Gaussian. For the following calculations, we assume a \OIL[1304] fraction of 50\% to correct for the \SiII contamination, which corresponds -on average- to the results found in literature.

\subsection{Emission line measurements}
\label{sec:lines}

Here we present the results obtained by performing the fit of the lines of interest in Bz5.3 using the method described in Sect. \ref{sec:fit}. These are also summarized in Table \ref{tab:results}.

\begin{table}[]
    \centering
    \caption{Results from the emission line fitting procedure}
    \begin{tabular}{c|ccc}
    \hline
         Line & $\mu$Flux & FWHM & EW \\
         & [$\rm 10^{-18} ~erg ~s^{-1} ~cm^{-2}$] & [$\rm km ~s^{-1}$] & [$\AA$]  \\
         \hline
         \hline
         \OIL[1304]$^\dagger$ & $3.2_{-1.0}^{+1.4}$ & -- & $15_{-6}^{+8}$ \\
         \OIL[7774] & $<0.4$ & -- & -- \\
         \OIL[7790] & $<0.5$ & -- & -- \\
         \OIL[8446] & $1.86\pm 0.15$ & $940_{-100}^{+50}$ & $25\pm 2$\\
         \CaL[8498] & $0.39_{-0.11}^{+0.13}$ & -- & $5.7\pm 1.5$\\
         \CaL[8542] & $0.40\pm 0.11$ & -- & $5.4 \pm 1.5$\\
         \CaL[8662] & $0.34_{-0.14}^{+0.16}$ & -- & $4.6\pm 2 $\\
         \hline
         \FeL[4278] & $0.609\pm 0.108$ & -- & $9 \pm 2$ \\
         \FeL[5165] & $0.588_{-0.091}^{+0.094}$ & -- & $9 \pm 1$ \\
         \FeL[5274] & $0.528_{-0.079}^{+0.083}$ & -- & $7\pm 1$ \\ 
         \FeL[5335] & $0.313\pm 0.077$ & -- & $4 \pm 1$\\
         \MgII & $1.98\pm 0.34$ & -- & $24\pm 4$\\
         
         \hline
         \hline
         
    \end{tabular}
    \label{tab:results}
    \flushleft
     \footnotesize {{\bf Notes.} Fluxes reported here are not corrected for magnification. $\dagger$ Results reported here do not consider the possible contamination from \SiII. }
    
\end{table}

\vspace{0.2cm}

\noindent \textbf{\OIall.} 
We clearly detect both the \OILs[1304] bump at S/N~$= 5$, and \OILs[8446] at S/N$= 9.6$, while \OILs[7774] and \OILs[7790] remain undetected. Given that the prism spectral resolution (R) increases by a factor of 10 between $\lambda_{\rm rest}=1304 ~\AA$ and $\lambda_{\rm rest}=8446 ~\AA$, the \OILs[1304] bump remains unresolved, while for \OILs[8446] we measure a broadening of ${\rm FWHM}=940_{-100}^{+50}\rm ~km ~s^{-1}$. The measured observed fluxes are reported in Table \ref{tab:results}.
Given these estimates, we derive the following flux ratios for \OILs[7774], \OILs[7790] over \OILs[8446]: $r_{7774/8446}<0.23$, $r_{7790/8446}<0.28$. To estimate the upper limit on flux ratio between \OILs[1304] and \OILs[8446] we assuming a fraction of \OIL[1304] of 100\%, implying $r_{1304/8446}\leq 1.7_{-0.6}^{+0.8}$. If assuming the contamination of \SiII $\lambda1309$ on the \OIL[1304] to be of order 50\% (see Sect. \ref{sec:bump} for discussion), the flux ratio becomes $r_{1304/8446, \rm corr}=0.9\pm 0.3$. These correspond to photon flux ratios of ${\rm ROI_{UV}}<0.27_{-0.09}^{+0.12}$, and ${\rm ROI_{\rm UV,corr}}\leq 0.13_{-0.04}^{+0.06}$, respectively\footnote{The photon flux ratio (region of interest or ROI) between two emission lines, at $\lambda_i$ and $\lambda_j$, is defined as ${\rm ROI_{ij}}=n(\lambda_i)/n(\lambda_j)$, where $n(\lambda_i)=F_\lambda^i/E_i ~[photons \rm ~s^{-1} ~cm^{-2}]$ is the photon flux of the line $i$ with flux $F_\lambda^i ~[\rm erg ~s^{-1} ~cm^{-2}]$ and with energy of the emitting photons $E_i=hc/\lambda_i ~[\rm erg]$.}.

\vspace{0.2cm}
\noindent \textbf{\CaT triplet.} Each line in the calcium triplet is detected with ${\rm S/N}=5,5,4$, respectively. The measured observed fluxes are reported in Table \ref{tab:results}.  
 These correspond to relative ratios of $1:1.0_{-0.3}^{+0.5}:0.87_{-0.36}^{+0.47}$ for \OILs[8498]:\OILs[8542]:\OILs[8662] w.r.t \OILs[8498]. This is close to the ratios derived for I~Zw~1 by \citet{rudy2000}. The photon flux ratio for the \Ca multiplet and the \OIL (ROI$_{\Ca}$), assuming a representative wavelength of $8579\AA$, is ${\rm ROI}_{\Ca } = 0.62_{-0.12}^{+0.16}$.

\vspace{0.1cm}
\noindent \textbf{\Fe emission lines.} As shown in the right panel of Figure \ref{fig:zoom-spec-Fe2}, in the UV the Vestergaard's template seems to properly model the blend of \MgII and \Fe at $\lambda_{\rm rest}=2800 \AA$, and it reproduces the spectral shape at $2200 \AA \lesssim \lambda_{\rm rest}\lesssim 2600 \AA$. However, the observed flux intensity of the \Fe lines in the $2200 \AA \lesssim \lambda_{\rm rest}\lesssim 2600 \AA$ regime is not well described by the template. This discrepancy is partially solved when varying the flux ratios of the \Fe emission in $2200 \AA \lesssim \lambda_{\rm rest}\lesssim 2600 \AA$ compared to the \Fe emission around \MgII (see Appendix \ref{app:fe-fit} and \citealt{labbe2024}). The variation of intrinsic flux ratios implies that the observed gas conditions differ from those assumed in the template model \citep{vestergaard2001}, but it is difficult to draw strong conclusions based on the poor resolution of the available data. 

 Given the known discrepancies between iron templates, we also tried to fit \Fe in the UV using the \Fe template from \citet{tsuzuki2006}. Results are reported in detail in Appendix \ref{app:fe-fit}. Unlike the Vestergaard's template where \MgII is masked, \citet{tsuzuki2006} also model the \Fe emission beneath and blended with \MgII. Unfortunately, the low spectral resolution of the data prevents us from appreciating these differences between the templates. Indeed, the quality of the fit using Vestergaard's or Tsuzuki's template is the same.

 Alternatively, we modeled the UV \Fe emission with HOMERUN \citep{marconi2024}, a multi-cloud photoionization framework that fits observed line fluxes as a non-negative combination of \textsc{cloudy} models (see Appendix \ref{app:fe-fit} for details). The resulting \Fe template (no turbulence) reproduces the emission around \MgII but fails to match the $\sim2500\,\AA$ bump; single-cloud models with turbulence recover this feature, but disagree with constraints from other lines. This tension points to including turbulence in HOMERUN, testing ad-hoc ionizing continua for AGN LRDs, and ultimately doing full-spectrum fitting of the Fe II pseudo-continuum — steps we leave for future work.

In the optical, the \Fe lines are well modeled by a single Gaussian, but are all unresolved at the spectral resolution of the prism (see left panel of Figure \ref{fig:zoom-spec-Fe2}).

\section{Discussion}
\label{sec:disc}

\subsection{Evidence for the broad-line region stratification}
\label{sec:Fe}
The detection of iron emission in the UV and in the optical gives us unambiguous evidence of an AGN (see Sect. \ref{sec:intro}). This is further corroborated by the detection of the \OI and \Ca lines and the broad \Halpha emission. Although we are unable to spectrally resolve ---and thus model in detail --- the iron multiplets in the UV, the detection of the iron signatures across the spectrum supports a scenario in which emission lines throughout the spectrum arise from the BLR, photoionized by the central BH \citep[see also discussion in][]{labbe2024}. 

In cases like Bz5.3 with poor spectral resolution, there is a current debate whether these features arises from the broad emitting region or the narrower one, i.e. if these are broad permitted emission lines \citep{labbe2024, kovacevic2025}, or narrow forbidden ones \citep{lin2025}. 

Following the first interpretation, in the optical we detect two categories of \Fe emission lines. \FeL[4278] is classified as an `inconsistent' \Fe line, while \FeL[5165], \FeL[5274] and \FeL[5335] as `consistent' ones. The appearance of lines that should be negligible based on their transition probabilities, namely `inconsistent' lines, has been recently studied and classified by \citet{kovacevic2025}. Based on the reported classification,  \FeL[4278] belongs to the P+ group of `inconsistent' lines. These are thought to arise either from lines from higher energy levels at similar wavelengths overlapping with these weak lines or from the same levels as consistent lines but amplified by unknown atomic processes. The presence and strength of `inconsistent' emission lines is known to increase for FWHM$_{\Fe, \rm opt}<5000 ~\rm km ~s^{-1}$. In the case of Bz5.3, the prism resolution limits the measured FWHM to be $\lesssim 1800 ~\rm km ~s^{-1}$ in the optical and $\lesssim 3000 ~\rm km ~s^{-1}$ in the UV, consistent with this trend. The diversity in both intensity, and possibly widths, among the consistent and inconsistent lines suggests that they originate from regions with varying physical conditions and distances from the BH, indicating a stratification of the \Fe emitting regions, and thus of the BLR \citep{kovacevic2025}.  

The more stringent constraint on their line widths (FWHM$_{\Fe, \rm opt}\lesssim 1800 ~\rm km ~s^{-1}$) favours the second interpretation for the nature of the optical iron features. Indeed, \Fef have been recently detected and modeled in two local LRDs by \citet{lin2025}. Based on the comparison with photoionization models and the FWHM of \Fef ($\sim 200-300 ~\rm km ~s^{-1}$), they propose the \Fef-emitting region to lie right at the edge of the BLR, supporting the view of a stratified inner region, and just outside the cool gas envelope. For Bz5.3 the presence of the cool-gas envelope can be supported by the evidence of the Balmer break, given that Balmer absorption features, if present, are not detectable at the current resolution. 

 Finally, the difference in line widths between LILs and Balmer lines further supports the picture of a stratified BLR, although the resolution of the prism can only set upper limits for most LIL line widths. Specifically, for the Balmer lines, we derived FWHM$_{\Halpha}=4600\pm 50 ~\rm km ~s^{-1}$ and FWHM$_{\Hbeta}=5100\pm 500 ~\rm km ~s^{-1}$, while for the LILs we reported FWHM$_{\Fe, \rm UV}\lesssim 3000 ~\rm km ~s^{-1}$, FWHM$_{\Fe, \rm opt}\lesssim 1800 ~\rm km ~s^{-1}$, FWHM$_{\OIL}=940_{-100}^{+50} ~\rm km ~s^{-1}$, FWHM$_{\Ca}\lesssim 900 ~\rm km ~s^{-1}$. LIL line widths differ by 2-3 thousands $~\rm km ~s^{-1}$ from the Balmer lines, implying that the LIL-emitting region is farther away from the central source in the BLR.

As briefly stated in Sect. \ref{sec:intro}, there seems to be a link between BLR tracers and LRDs, possibly due to BH properties or geometric effects. We note that both LRDs with robust \Fe detection, Bz5.3 and A2744-45924, have a low Eddington ratio ($\lambda_{\rm Edd, Bz5.3}=0.010_{-0.007}^{+0.020}$, and $\lambda_{\rm Edd, A2744}<0.03$, from \citealt{ji2025, furtak2024,juodzbalis2025}). However, if black-hole masses from single-epoch virial calibrations are to be trusted, this seems to be in contrast to many studies reporting a correlation between \Fe strength and Eddington ratio \citep{kovacevic2025, sulentic2000, shen2014,gaskell2022, martinez2021}. In contrast, if we believe mass estimates from the electron-scattering scenarios, then most AGN LRDs would be super-Eddington (or even highly super-Eddington, in the quasi-star model). As for the current state of the models and observations available, the origin of this link remains an open question.

\subsection{Dust attenuation and origin of UV continuum}
\label{sec:av}

According to the Grotrian diagram\footnote{A Grotrian diagram, or term diagram, shows the allowed electronic transitions between the energy levels of atoms. They can be used for one-electron and multi-electron atoms.} for \OI \citep[see e.g. Figure~3 in][]{grandi1980}, the ROI of \OILs[1304]/\OILs[8446] (ROI$_{\rm UV}$) must be equal to unity, in principle only decreasing when reddening is present. This is why the photon flux ratio between \OILs[1304] and \OILs[8446] has been used as reddening indicator \citep[e.g., ][]{rudy1989, laor1997, rodriguez2002, matsuoka2005}. These works also suggest that the ROI$_{\rm UV}$ can deviate from unity\footnote{Values of ROI$_{\rm UV}$ $= 1$ and $= 0.63$ corresponds to intrinsic flux ratios of 6.5 and 4, respectively.}, down to $0.63$ even in the absence of reddening. This could be  due to (i) Balmer continuum absorption of \OILs[1304] photons, or (ii) the production of \OILs[8446] by collisional excitation, or (iii) by collisional de-excitation destroying \OILs[1304] photons \citep{kwan1981, grandi1980}. For Bz5.3, we estimated an ROI$_{\rm UV}$ between 0.1 and $0.3$ (see Sect. \ref{sec:lines}) therefore, even if the \OILs[1304] is severely affected by destruction mechanisms or the \OILs[8446] is pumped by collisional excitation, these mechanisms are not sufficient to explain such a break of the one-to-one photon relation between \OILs[8446] and \OILs[1304] \citep[as also discussed in][]{rodriguez2002}. Since we do observe a Balmer break
in Bz5.3, and since these features have been associated with dense gas near AGN \citep{inayoshi2025, ji2025, naidu2025, degraaff2025}, a discussion of Balmer-continuum absorption is warranted. Comparing Bz5.3 to A2744-45924 (see Figure~\ref{fig:phot-spec}) reveals that our AGN has a weaker break and steeper UV slope. This implies that Balmer absorption near \OILs[1304] may not be dominant, since the weak break compounds the decreasing bound-free cross section
blueward of the Balmer limit. Independent evidence for weak neutral-gas absorption comes from the sharpness of the Ly$\alpha$ break in Bz5.3, suggesting that the UV photons propagate through a path largely free of neutral hydrogen, which would otherwise leave a spectral-absorption imprint, as seen in the smooth Ly$\alpha$ break of A2744-45924. Thus, we favour dust playing a major role in altering the observed \OI ratio. 

We derive $E(B-V)$ and the dust attenuation ($A_V$) considering the observed ROI$_{\rm UV}$ derived in Sect. \ref{sec:lines}, both the intrinsic ROI$_{\rm UV}$ presented above, and assuming the Calzetti law for the dust attenuation curve \citep{calzetti1996}. Results are reported in Table \ref{tab:attenuation}.  We also consider the limit case of having a fraction of \OIL[1304] of 20\%. Our findings point toward relatively high values of dust attenuation in the BLR, ranging from 0.4 to 1.4. 

To assess the origin of the continuum under the \OIL[1304] line, we use the range of derived dust attenuation to correct the observed UV $\beta$ slope, $\beta_{\rm UV}=-1.6\pm 0.1$. Assuming $A_V \in [0.4,1.0]$, we obtain intrinsic UV beta slopes of $\beta_{\rm UV, int}=[-2.8,-2.0]$. This is broadly consistent with a standard accretion disk \citep[i.e. $\beta=-2.3$, see][]{shakura1976}, and towards the steepest end with extreme star formation. Similar steep intrinsic UV slopes  have been recently found both in the local LRDs at $z\sim 0.1-0.2$ \citep{lin2025}, and in LRDs at $z>4$ \citep[see e.g.][]{labbe2024, tripodi2024}. These works favor the interpretation of the UV being dominated by AGN light, also motivated by the high EW of the UV emission lines.

\begin{table*}[]
    \centering
    \caption{$E(B-V)$ and dust attenuation from the \OI lines}
    \begin{tabular}{c|c|c|c|c|c|c}
         & \multicolumn{3}{c|}{ROI$_{\rm UV, int}=1.0$} & \multicolumn{3}{c}{ROI$_{\rm UV, int}=0.63$}\\
         \hline
         & no \SiII & 50\% \SiII (fiducial) & 80\% \SiII & no \SiII & 50\% \SiII (fiducial) &  80\% \SiII \\
         \hline
         $E(B-V)$ & $0.16 \pm 0.04$ & $0.24 \pm 0.04$ & $0.36 \pm 0.05$ & $0.10\pm 0.04$ & $0.19 \pm 0.04$ & $0.30 \pm 0.05$\\
         
       $A_V$  & $0.65\pm 0.20$& $1.0\pm0.2$ & $1.4 \pm 0.2$ & $0.42 \pm 0.20$& $0.80 \pm 0.20$ & $1.2 \pm 0.2$\\
       \hline
       \hline
    \end{tabular}
    \label{tab:attenuation}
    \flushleft
    \footnotesize {{\bf Notes}. The columns specify the assumptions made to compute the values of dust attenuation.}
\end{table*}

\subsection{Insights from the \Ca transitions}

Comparisons of the \Ca triplet with \Hbeta and \Fe feature strengths could better constrain BLR properties and geometry. However, such analysis is hindered by: (1) the limited resolution of the prism, which prevents detailed modeling of UV \Fe emission; and (2) large uncertainties in de-reddened line fluxes due to the variable dust attenuation inferred from the \OIL[1304]/\OILs[8446] ratio. Estimating attenuation from the broad Balmer lines is even more uncertain due to blending with nearby features (e.g., \OIIIL, \OIIIL[4364], \Fe) at the current resolution.

The near-unity relative strengths of the \CaT lines suggest they are optically thick \citep{ferland1989}, consistent with the redshift in their centroids (Figure~\ref{fig:zoom-spec}), as expected from resonant scattering. Other \Ca lines—such as H, K, and the forbidden \OILs[7291] and \OILs[7324]~\AA—are typically absent in AGNs \citep{ferland1989} and are also undetected in Bz5.3. The \CaT/\OIL flux ratio varies across AGNs but is generally near unity \citep[minimum $0.35 \pm 0.19$;][]{ferland1989}; for Bz5.3, we find $0.61^{+0.15}_{-0.12}$, consistent with this range.

\subsection{Excitation mechanisms in the BLR}
\label{sec:ext-mec}

If Ly$\beta$ fluorescence is the dominant excitation mechanism, the \OI\ transitions at \OILs[1304], \OILs[8446], and \OILs[11287] should appear in a 1:1:1 photon flux ratio. However, this ratio can be altered by three competing processes: recombination, collisional excitation, and continuum fluorescence. The presence or absence of auxiliary \OI\ lines such as \OILs[7774] and \OILs[7990] provides diagnostic power to identify the dominant mechanisms \citep[see Sect.~\ref{sec:intro};][]{grandi1980, landt2008, matsuoka2005, matsuoka2007, matsuoka2008}. 

Ly$\beta$ fluorescence and collisional excitation are the main sources of \OIL\ emission in the BLR. \citet{grandi1980} first identified the absence of a narrow \OIL component — supporting a BLR origin — and proposed Ly$\beta$ fluorescence as the dominant mechanism, later confirmed by the 1:1:1 line ratio in I~Zw~1 \citep{rudy1989}. Subsequent studies showed collisional excitation is also significant \citep{rodriguez2002, landt2008}, while continuum fluorescence plays a minor role. Photoionization models by \citet{matsuoka2007} showed both mechanisms operate efficiently in gas with $n_{\rm H} \sim 10^{11.5}~\rm cm^{-3}$ and $\log(U) \sim -2.5$.

In Bz5.3, we clearly detect \OIL[8446], \CaT triplet and we have upper limits on \OIL[7774] and \OIL[7990]. Consistent with a BLR origin, \OIL[8446] exhibits broad emission (see Table \ref{tab:results}), though it is several thousand $\rm km~s^{-1}$ narrower than the broad Balmer lines—supporting the notion of a stratified BLR.  
In agreement with the aforementioned studies in local AGNs, for Bz5.3, Ly$\beta$ fluorescence is the main production mechanism of LILs based on the following evidence. (1) The high equivalent width of \OIL[8446] can be explained only in the presence of Ly$\beta$ pumping the production of \OIL[8446] photons. Indeed, it has been shown that in the presence of mechanisms hampering the Ly$\beta$ fluorescence efficiency, e.g. turbulent gas, the EW of \OIL[8446] decreases to $<10\AA$ \citep{matsuoka2007, bottorff2000}. (2) The weakness of the \OIL[7774],\OILs[7990] emission lines compared to \OIL. Indeed their flux ratios are expected to be $>1$ for pure recombination, while \OIL[7774],\OILs[7990] are just a few percents of \OIL ($<10-20\%$) in case of a combination of Ly$\beta$ fluorescence and collisional excitation \citep{matsuoka2007}.

Indeed, we also observe the presence of other competing mechanisms given that the \OIL[1304]/\OILs[8446] photon flux ratio is much lower than unity. We can rule out recombination as a player in fostering the production of LILs, since \OIL[7774]/\OILs[8446]$<0.23$ (\OIL[7774]/\OILs[8446]$\sim 1.1$ for recombination). Instead, this flux ratio is close to the expected one for collisional excitation \citep[0.3, see][]{matsuoka2007}. In contrast, the  \OIL[7990]/\OILs[8446] ratio of $<0.28$ is rather uninformative, since it is expected to be $\sim 0.05$ for continuum fluorescence \citep{grandi1980}. 

We can rule out continuum fluorescence given that EW(\OIL[8446])$=24\pm 2 ~\AA$, while EW(\OIL[8446])$\leq 1 ~\AA$ is expected \citep{matsuoka2007}. This high equivalent width also supports the role of collisional excitation, which is more effective for higher EW(\OIL[8446]) that implies higher densities \citep{matsuoka2007}.

The contribution from scattered light is also ruled out, since we would expect much higher \OIL[1304]/\OILs[8446] ratio than observed, roughly proportional to the scattered fraction.

The complexity of the BLR in LRD is highlighted by comparing the results for the EW(\OIL[8446]) and ROI$_{\Ca}$ with simple photo-ionization models. In particular, we use the models provided in \citet{matsuoka2007}, who performed a similar analysis for QSOs at $z<1$. The BLR gas in their work was modeled to have a constant hydrogen density and exposed to the ionizing continuum radiation with varying photon flux, $\Phi$, connected to the ionization parameter as $U\equiv \Phi/(n_{\rm H}c)$ where $c$ is the speed of light. They performed calculations with $(n_{\rm H}, U)$ sets in a range of $10^7\leq n_{\rm H}\leq 10^{14} \rm ~cm^{-3}$, and $10^{-5}\leq U\leq 10^0$, stepped by 0.5 dex. They have a BLR gas with $N_{\rm H}=10^{23} \rm ~cm^{-2}$ and $v_{\rm turb}=0 ~\rm km ~s^{-1}$, where $N_{\rm H}$ is the column density and $v_{\rm turb}$ is the microturbulent velocity. We do not consider other models presented in their work with higher $v_{\rm turb}$ values, since they imply EW(\OIL[8446])$\leq 10 ~\AA$, in contrast with our observations. Therefore, based on their `standard' model and the observed EW(\OIL[8446]) and ROI$_{\Ca}$ for Bz5.3, we find that $-4<\log(U)<-3.5$, and $n_{\rm H}\sim 10^{9.5} ~\rm cm^{-3}$ \citep[see Figure~7 in][]{matsuoka2007}. However, other models favour higher densities and ionization to reproduce \Fe emission in local AGNs \citep[$\log(n_{\rm H}/\rm cm^{-3})\sim 10-11$, $\log(U)\sim -2.5$, e.g.][]{lin2025}. A step forward would be to generate tailored models to reproduce the observed complex stratification observed in the BLR taking into account constraints from \Fe emission, which is however beyond the scope of this Letter \citep[see e.g., ][]{lin2025, panda2020, santos2024}.

\section{Summary and Conclusions}
\label{sec:concl}

In this work, we present a spectroscopic study of BLR LIL diagnostics in Bz5.3, a $z = 5.3$ AGN host galaxy, based on \textit{JWST}/NIRCam and NIRSpec observations covering rest-frame UV to optical wavelengths. Our main conclusions are as follows:

\begin{itemize}
    \item The detection of \Fe, \OILs[1304] bump, broad \OI, and \Ca triplet emission lines supports the presence of a BLR and thus an AGN in Bz5.3, already suggested by evidence of broad Balmer emission.

    \item We report the first detection of the $\lambda1304$ bump at $z>5$, a key spectral signature of the BLR previously seen only in local AGN. This detection extend the study of local diagnostics of the BLR physical conditions to $z>5$.

    \item While the iron features in Bz5.3 may arise from either broad permitted or narrow forbidden transitions, the diversity in line strengths and widths—along with their consistency with recent models—supports the presence of physically distinct emitting regions. Regardless of the interpretation, the observed iron emission points to a stratified BLR.

    \item Excitation diagnostics suggest that Ly$\beta$ fluorescence is the primary mechanism powering the \OI lines, particularly \OIL[8446], with collisional excitation also contributing. Recombination and continuum fluorescence are disfavored based on upper limits on \OIL[7774] and \OIL[7990], and the large equivalent width of \OIL[8446] ($24 \pm 2$ \AA).

    \item The photon flux ratio \OIL[1304]/$\lambda8446$ is significantly below unity (0.1–0.3), pointing to substantial dust attenuation within the BLR. We derive $A_V \sim 0.4$--$1.4$ based on these ratios. We disfavour bound-free absorption since the weak break compounds the decreasing bound-free cross section blue-ward of the Balmer limit.

    \item The \Ca triplet shows nearly equal line ratios, indicating optically thick conditions. This observation is consistent with an observed redshift of the line centroid, as expected from resonant scattering. The measured \Ca/\OIL[8446] flux ratio ($\sim 0.6$) is consistent with low-redshift AGNs, further supporting a BLR origin for these lines.

    \item  Comparison with photoionization models suggests that the observed \OIL[8446] and ROI$_{\Ca}$ features in Bz5.3 are best reproduced by a low-ionization, high-density BLR component, distinct from the conditions typically required for iron emission. This apparent tension further supports a stratified BLR structure, with multiple gas phases coexisting under different physical conditions—underscoring the need for tailored models to capture this complexity.

\end{itemize}

A detailed study of the BLR through LIL diagnostics such as \Fe, \OI and \Ca, have usually been limited to the low-z Universe ($z<1$), mainly due to a combination of poor resolution and sensitivity. This is especially true for \OIL[1304], which is usually faint and often blended with Ly$\alpha$ emission or hidden by Ly$\alpha$ damping wings \citep[see e.g.,][]{labbe2024}. Taken together, our findings provide a rare and detailed glimpse into BLR conditions in LRDs at early times, demonstrating the power of \textit{JWST} spectroscopy in probing AGN physics in the early Universe. Nonetheless, the limited resolution of the NIRSpec prism restricts a robust analysis of LILs and the BLR structure, motivating follow-up observations with higher-resolution spectroscopy.

While extending this analysis to a large population of high-redshift galaxies remains challenging — particularly for the faint and blended \OIL[1304] bump — targeted, deep observation of carefully selected sources offers a promising path forward for probing the structure and physics of the BLR in the early Universe. In parallel, the recent discovery of local LRD analogs provides a valuable laboratory for investigating BLR properties in greater detail. Tailored UV spectroscopic follow-up with instruments like HST/COS for local LRDs and JWST/NIRSpec gratings for high-z ones will be essential for bridging the gap between low- and high-redshift LRDs, enabling a more complete picture of BLR excitation mechanisms and evolution over cosmic time.

\bigskip

 Data Avalability: all the {\it JWST} data used in this paper can be found in MAST: \dataset[10.17909/2hxk-5h89]{http://dx.doi.org/10.17909/2hxk-5h89}.

\begin{acknowledgements}

We thank Dr. Marianne Vestergaard and Dr. Yoshiki Matsuoka for kindly sharing the iron templates used in this work and their insightful suggestions. The authors also thank Dr. Cupani and Dr. Calderone for their useful suggestions on the use of iron templates. RT, MB, NM, VM, AH, GR, JJ, and GF acknowledge support from the ERC Grant FIRSTLIGHT and from the Slovenian national research agency ARIS through grants N1-0238 and P1-0188. GR and MB acknowledge support from the European Space Agency through Prodex Experiment Arrangement No. 4000146646. This research was enabled by grant 18JWST-GTO1 from the Canadian Space Agency, and Discovery Grants from the Natural Sciences and Engineering Research Council of Canada to MS, RA, and AM. This research used the Canadian Advanced Network For Astronomy Research (CANFAR) operated in partnership by the Canadian Astronomy Data Centre and The Digital Research Alliance of Canada with support from the National Research Council of Canada, the Canadian Space Agency, CANARIE and the Canadian Foundation for Innovation. This work is based on observations made with the NASA/ESA/CSA James Webb Space Telescope. The data were obtained from the Mikulski Archive for Space Telescopes at the Space Telescope Science Institute, which is operated by the Association of Universities for Research in Astronomy, Inc., under NASA contract NAS 5-03127 for JWST. These observations are associated with program \#4598.  Support for program \#4598 was provided by NASA through a grant from the Space Telescope Science Institute, which is operated by the Association of Universities for Research in Astronomy, Inc., under NASA contract NAS 5-03127. BL is supported by the international Gemini Observatory, a program of NSF NOIRLab, which is managed by the Association of Universities for Research in Astronomy (AURA) under a cooperative agreement with the U.S. National Science Foundation, on behalf of the Gemini partnership of Argentina, Brazil, Canada, Chile, the Republic of Korea, and the United States of America.
    
\end{acknowledgements}

\appendix

\section{\texttt{Galfit} fit}
\label{app:galfit}
\begin{figure}
    \centering
    \includegraphics[width=0.45\linewidth]{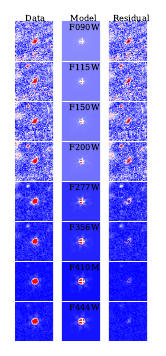}
    \includegraphics[width=0.45\linewidth]{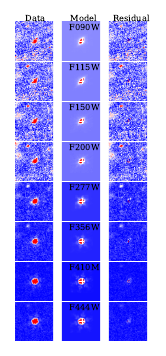}
    \caption{Results of modeling Bz5.3 with \texttt{Galfit}. Left panels: data and results assuming a single point source as model for each filter. The white cross indicates the centroid of the model. Right panels: data and results assuming as model a combination of a point source and a Sérsic profile for each filter with $\lambda<4~\mu m$, a single point source otherwise.}
    \label{fig:galfit}
\end{figure}

In the left panel of Figure~\ref{fig:galfit}, we show the results of the \texttt{Galfit} modeling of Bz5.3 with a point source. We find strong residuals in all filters except F410M and F444W, where the source is unresolved and fully consistent with a point source. Therefore, we add an additional component for all the filters at $\lambda<4~\mu m$ in the form of a Sérsic profile with varying index $n$ and half-light radius $R_e$. As shown in the right panel of Figure~\ref{fig:galfit}, this setup produces goods results with few residuals.

\section{Details on the spectral fitting procedure}
\label{app:fit}

Regarding the fit of the spectral features reported in Sect. \ref{sec:fit}, we explore the parameter space with a Markov chain Monte Carlo (MCMC) algorithm implemented in the \texttt{EMCEE} package, assuming uniform priors for slopes, normalizations, and FWHM of the Gaussian. Priors on the FWHM of narrow lines are tight, taking into account the known varying resolution with wavelength of the prism. We consider $\rm FWHM_{prior} \in [0.8,1.2]$ spectral resolution element, whose size vary with wavelength. The prior on the peak wavelength is Gaussian, centered on $z_{\rm source}=5.2907$, and with $\sigma=10 \AA$, corresponding to the best spectral resolution achieved in the spectrum at the source's redshift. We consider $10$ walkers per parameter and 2500 trials, with a typical burn-in phase of $\sim 300$ trials. 

We compute the integrated line fluxes by integrating the best-fitting functions for each emission line, which are quoted as the 50th percentile, and we derive the error from the 16th and 84th percentiles. Upper or lower limits are given at $3\sigma$. When measuring line widths, an additional complication is given by the instrumental broadening effect by the detector line spread function (LSF), which can be approximated as a Gaussian profile of a certain FWHM \citep{jakobsen2022}. Therefore, we report de-convolved FWHM computed via $\rm FWHM_{dec}=\sqrt{\rm FWHM_{obs}^2-FWHM_{LSF}^2}$ relation, where $\rm FWHM_{obs}$ and $\rm FWHM_{LSF}$ are the observed and instrumental FWHM (publicly available in the JWST website), respectively. We mention that LSF correction are usually negligible for broad lines in the rest-frame optical regime, e.g. $>10$ times wider than the $\rm FWHM_{LSF}$. FWHM below the spectral resolution of the instrument are not reported. We measure the EW of emission lines for each trial in the chain. Then, we derive the best-fitting value from the 50\textsuperscript{th} percentile of the EW's chain, and the error on EW from the 16\textsuperscript{th} and 84\textsuperscript{th} percentiles.

\section{Alternative \Fe and \MgII fits}
\label{app:fe-fit}

\begin{figure}[h]
    \centering
    \includegraphics[width=0.9\linewidth]{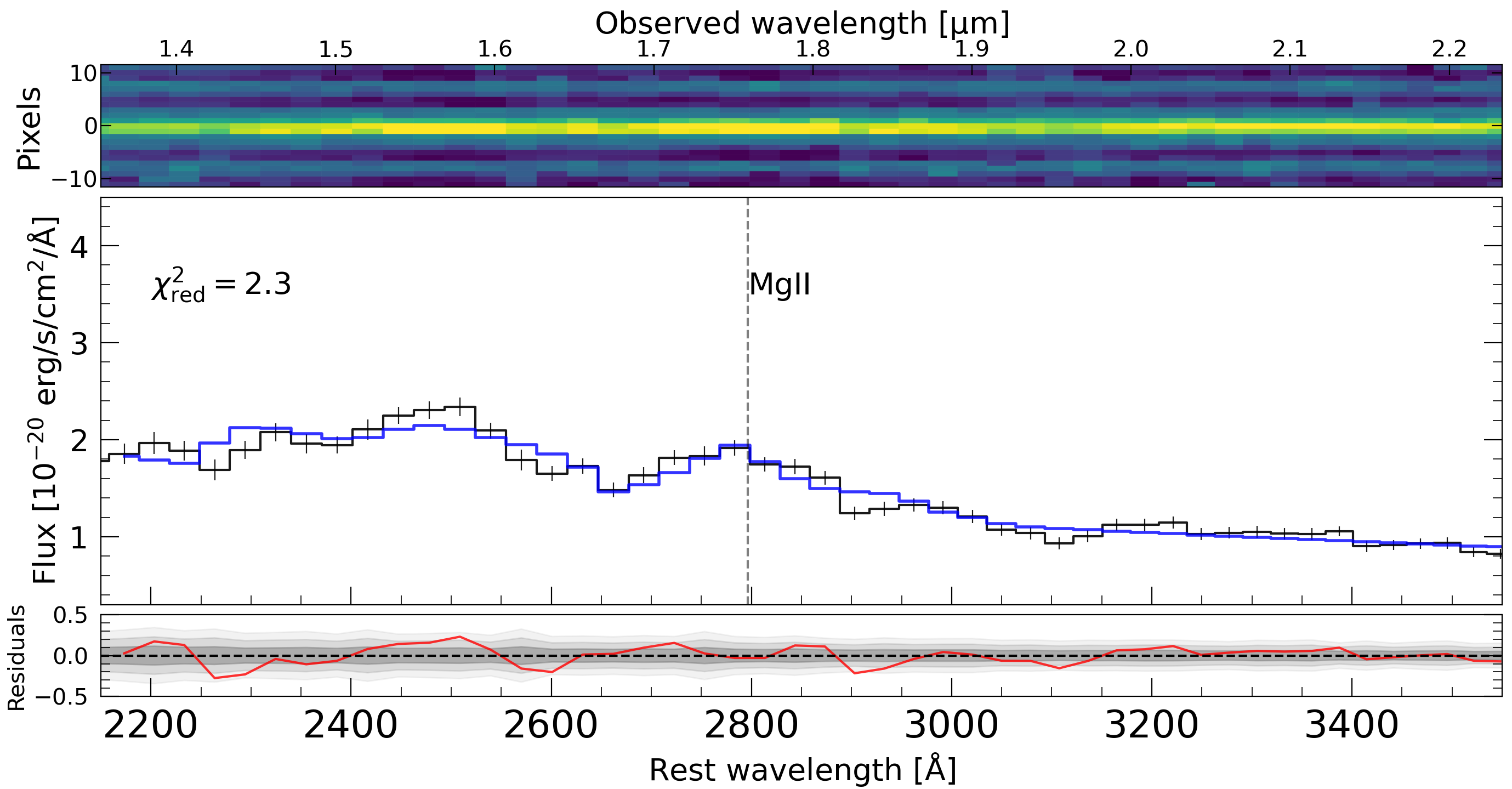}
    \caption{Zoom-in of the 2D and 1D spectra of Bz5.3 on the \Fe complex in the UV and \MgII emission line. The observed spectrum is shown in black, while the best-fitting model is shown in blue. Residuals are shown at the bottom of each panel, where the shaded dark gray area represents the 1$\sigma$ noise level. 2 and 3$\sigma$ noise levels are reported in the residual as shaded lighter gray area.}
    \label{fig:zoom-spec-Fe2-alt}
\end{figure}

\begin{figure}[h]
    \centering
    \includegraphics[width=0.9\linewidth]{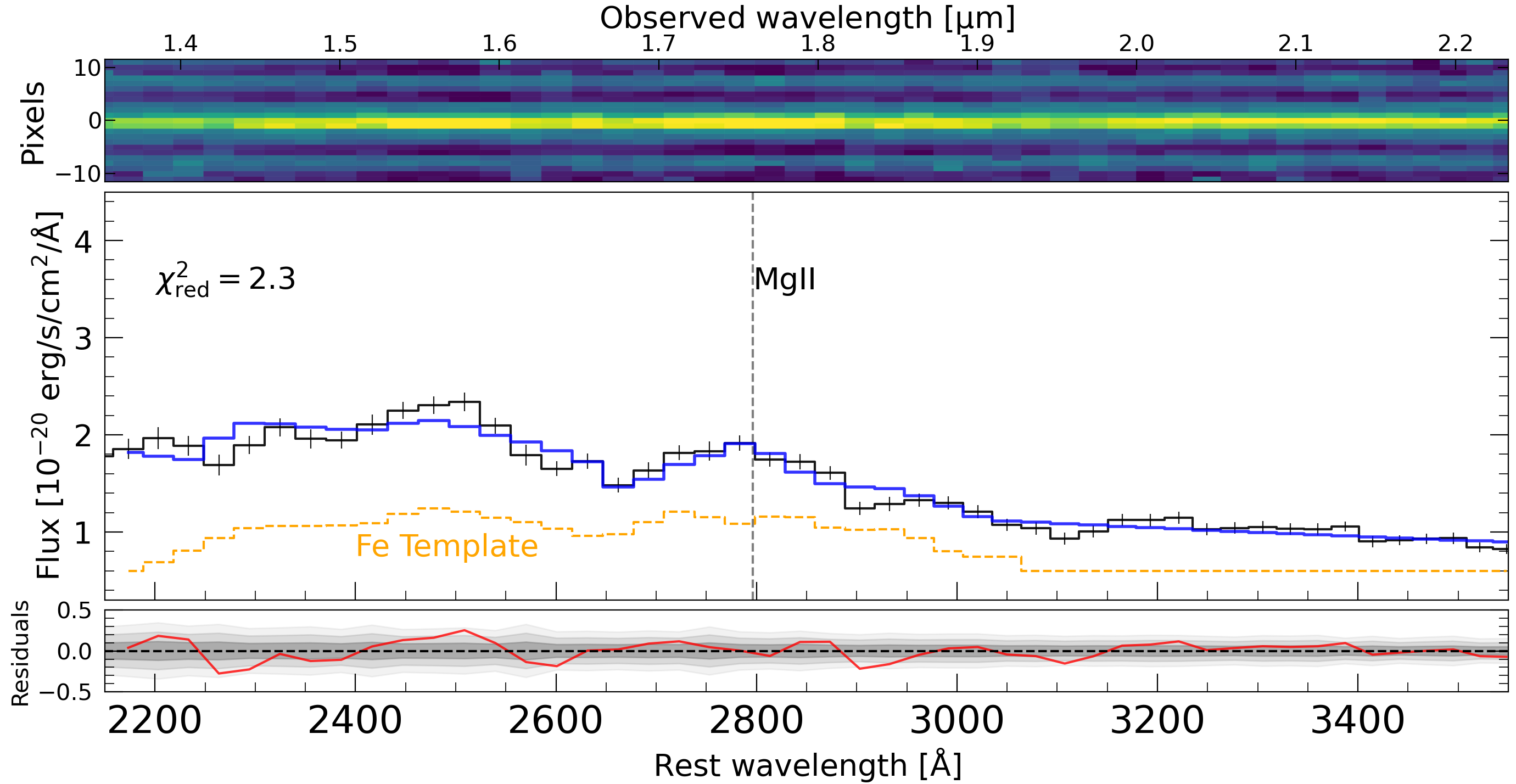}
    \caption{Same as Figure \ref{fig:zoom-spec-Fe2-alt} but for Tsuzuki's template (shown in yellow).}
    \label{fig:Fe2-tsuzuki}
\end{figure}

\begin{figure}[h]
    \centering
    \includegraphics[width=0.9\linewidth]{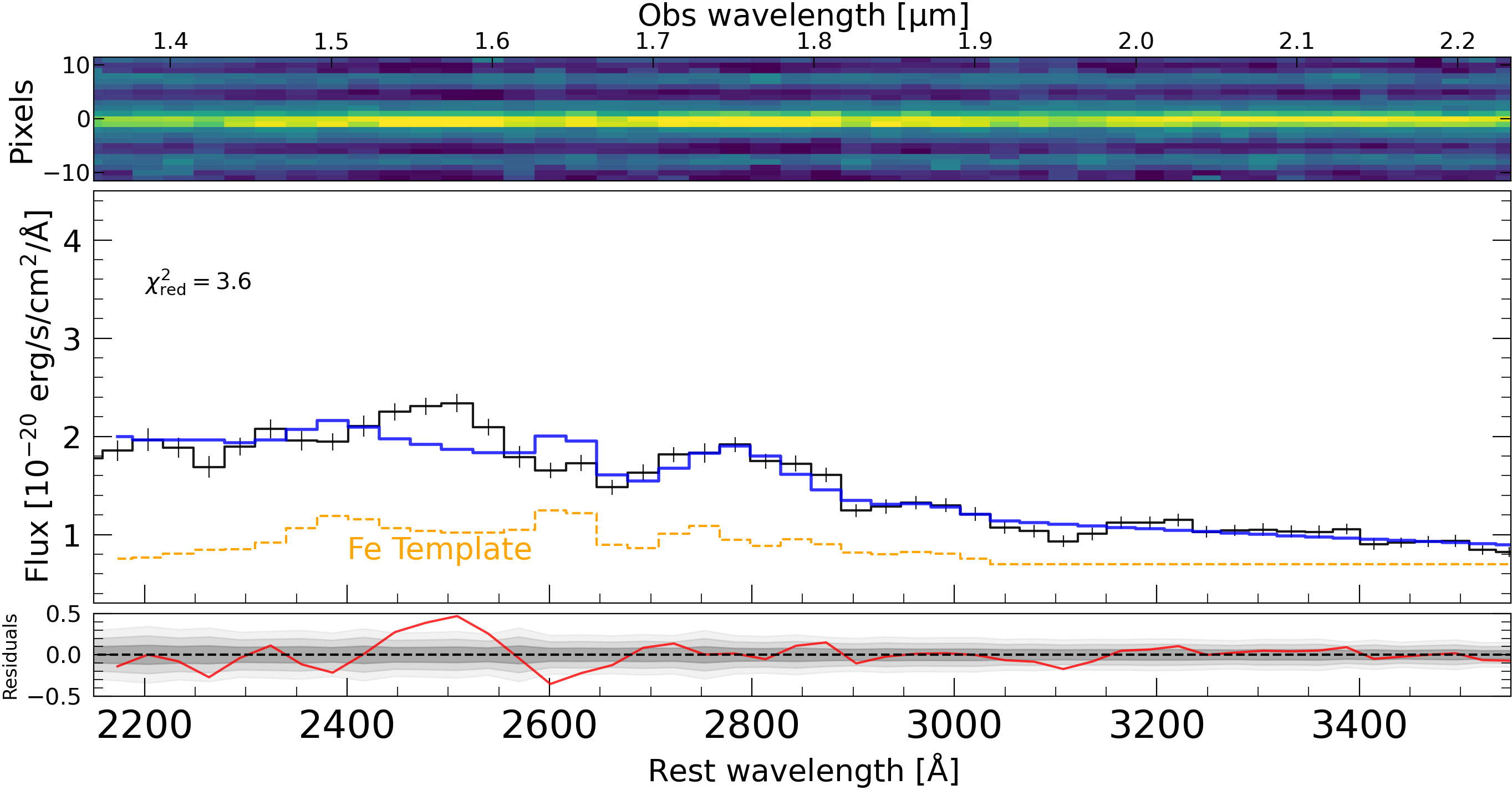}
    \caption{Same as Figure \ref{fig:zoom-spec-Fe2-alt} but for the HOMERUN template (shown in yellow).}
    \label{fig:Fe2-cloudy}
\end{figure}

As shown in the bottom panel of Figure~\ref{fig:zoom-spec-Fe2} and in Sect. \ref{sec:lines}, the template from \citet{vestergaard2001} are able to reproduce the overall shape of the \Fe emission in $2200 \AA\lesssim \lambda_{\rm rest} \lesssim 2600 \AA$, however the flux ratios of the \Fe transitions assumed in the template differ from the observed ones in Bz5.3. Therefore, to allow more flexibility in the model, we splitted the template into two sections separated at $\lambda_{\rm rest}=2700 \AA$, isolating the \MgII emission region from the rest of the \Fe pseudo-continuum. We show the results in Figure~\ref{fig:zoom-spec-Fe2-alt}. The fit is now much closer to the observed data at $\lambda_{\rm rest}<2700 \AA$, as shown by the lower reduced $\chi^2$ value compared to the previous fit. This implies that a proper modeling of the \Fe emission would require higher resolution data to better resolve the pseudo-continuum followed by templates with different assumptions for the gas conditions. 

 Alternatively, given the known discrepancies with \Fe templates, we run an analogous fit using the template from \citet{tsuzuki2006}. As shown in Figure \ref{fig:Fe2-tsuzuki}, we obtained a similar result as when using Vestergaard's template. Indeed, the low resolution of the data in the UV regime prevents us from distinguishing between different models of \Fe emission.

Seeking a more reliable match to the UV \Fe emission, we adopted a tailored photoionization model for the observed data. Specifically, we modeled the emission lines listed in Table~1, together with \CIII, \CIV, and the broad components of \Halpha\ and \Hbeta, using HOMERUN \citep{marconi2024, ceci2025}. HOMERUN (Highly Optimized Multi-cloud Emission-line Ratios Using photoionizatioN; \citealt{marconi2024}) is a multi-cloud photoionization framework in which the observed emission lines are reproduced as a weighted linear combination of constant-density \textsc{cloudy} models, each characterized by a specific ionizing photon flux and density but sharing the same ionizing spectrum and chemical composition. The key innovation is that the weights are treated as free parameters and determined by fitting the observed line fluxes to the model predictions via non-negative least-squares minimization. From the HOMERUN best-fit model, we inferred the combination of gas densities and ionizing photon fluxes consistent with our observations (specifically,\footnote{The barred quantities are the harmonic means of the gas density and ionizing photon flux, respectively.} $\log(\bar{n})=10.9$, $\log(\bar{\phi})=17.7$). We then computed synthetic \Fe templates with \textsc{cloudy} v23.01 \citep{gunasekera2023}, adopting the table AGN ionizing continuum \citep{mathews1987} and matching the metallicity to the HOMERUN best-fit value ($12+\log(\mathrm{O/H})=7.8$). To ensure that all relevant atomic levels and transitions were included, we used the full \Fe model atom implemented in \textsc{cloudy}, based on the Stout database \citep{ferland1989} with updated atomic data from \citet{smyth2019}. No turbulence was included, consistent with the HOMERUN best-fit model. The resulting template and its fit to our data are shown in Fig.~\ref{fig:Fe2-cloudy}. While the new template reproduces the observed \Fe emission around \MgII well, it fails to recover the $\sim2500\,\AA$ bump, unlike the templates by \citep{vestergaard2001} and \citep{tsuzuki2006}. Given the strong sensitivity of \Fe emission to gas turbulence, we also tested simpler single-cloud models with non-zero turbulence, which reproduce the $\sim2500\,\AA$ bump but are inconsistent with the observed fluxes of other lines. A natural step forward would be to investigate the effects of including non-zero turbulence in HOMERUN, as well as building 'ad hoc' ionizing continuum shapes for AGN LRDs. Moreover, a proper treatment of \Fe pseudo-continuum would require a full-spectrum fitting, which is not implemented in the current version of HOMERUN. While it would be a natural extension of the code, this is beyond the scope of this paper.

\bibliography{biblio,lensing}{}
\bibliographystyle{aasjournalv7}

\end{document}